\documentclass[12pt]{article}
\usepackage{latexsym,amsfonts}%,maple2e}
\usepackage{color}
\usepackage{amsmath}
\usepackage{amssymb}
\usepackage{epsfig}
\usepackage{bm}
\begin{document}
\title{\bf \LARGE Fermions, Skyrmions and the 3-Sphere}
\author{Stephen W. Goatham{\thanks{E-mail: {\tt
	swg3@kent.ac.uk}}}\hspace{1.5mm} and Steffen
  Krusch\thanks{E-mail: {\tt S.Krusch@kent.ac.uk}}\\ \\[5pt]
{\normalsize {\sl School of Mathematics, Statistics and Actuarial
Science}}\\
{\normalsize {\sl University of Kent,
Canterbury CT2 7NF, United Kingdom}}
}

\date{August 14, 2009}
\maketitle

\begin{abstract}
This paper investigates a background charge one Skyrme field chirally
coupled to light fermions on the 3-sphere.  The Dirac equation for the
system commutes with a generalised angular momentum or grand spin. It can
be solved explicitly for a Skyrme configuration given by the hedgehog
form. The energy spectrum and degeneracies are derived for all values
of the grand spin. Solutions for non-zero grand spin are each
characterised by a set of four polynomials. The paper also discusses
the energy of the Dirac sea using zeta function regularization.
\end{abstract}

\section{Introduction}
The Skyrme model is a nonlinear $SU(2)$ field theory which gives a good
description of atomic nuclei and their low energy interactions
\cite{Skyrme:1961vq}.  In addition to the fundamental pion excitations
the theory also has topological soliton solutions, known as
Skyrmions. These are labelled by a topological
charge or generalised winding number $B$, which can be interpreted as
the baryon number of the configuration. On
quantization, Skyrmions are found to describe nuclei,
$\Delta$-resonance \cite{Adkins:1983ya} and also bound states of
nuclei, see \cite{Kopeliovich:1988np, Braaten:1988cc, Leese:1994hb,
  Irwin:1998bs, Krusch:2002by, Krusch:2005iq} for the
quantization of multi-Skyrmions and \cite{Battye:2006na, Manko:2007pr,
  Battye:2009ad} for recent quantitative 
predictions of the Skyrme model. It is well known
that Skyrmions can be quantized as fermions
\cite{Finkelstein:1968hy,Witten:1983tw}.
Therefore, when the Skyrme field is coupled to a fermion field, 
there are two different ways of describing fermions in the
same model.
The fermion field can then be thought of as light quarks in the
presence of
atomic nuclei, \cite{Balachandran:1998zq}. In the presence of a
Skyrme field the energy spectrum of the Dirac operator shows a
curious behaviour, namely, a mode crosses from the positive to the
negative spectrum as the coupling constant is increased,
\cite{Hiller:1986ry}. In a very similar model, Kahana and Ripka
calculate the baryon density in the one-loop approximation \cite{
Kahana:1984be} and the energy of the Dirac sea quarks
\cite{Ripka:1985am}. Recently, these calculations have been extended
to multi-Skyrmions, \cite{Komori:2003zn,Sawado:2002zm}.

Static field configurations in the original Skyrme model in flat space
are given by maps $\mathbb R^{3}\rightarrow SU(2)$. By using the
boundary condition for the Skyrme field
to unify the domain of such a map with infinity we make the domain
compact and equivalent to the 3-sphere $S^{3}$. 
If we also consider that $S^{3}$ is the group manifold of $SU(2)$,
we can see that the field configurations are topologically equivalent
to maps $S^{3}\rightarrow S^{3}$ and, because of this, the
model can be generalized to the base space being a sphere of radius
$L$, \cite {Manton:1987xt, Krusch:2000gb}. In the limit
$L\rightarrow \infty$, the original model is recovered. The advantage
of working on $S^3$ is that the Bogomoln'yi equation can be solved for
$B=1,$ and the solution is given by the identity map
\cite{Manton:1987xt}. This enables us to calculate the energy
spectrum and the corresponding fermion wave functions explicitly.

In \cite{Krusch:2003xh} a system of light fermions, on $S^{3}$,
coupled to a spherically symmetric background Skyrme field was studied
for grand spin $G=0$. In this paper we consider the general case where
the grand spin also takes positive integer values. The Dirac equation
on $S^3$ is derived in section \ref{Dirac} through the use of stereographic
coordinates. In section \ref{G=0}, the solution of the Dirac equation
for $G=0$ is reviewed.
In section \ref{G}, the correct ansatz for the spin-isospin
spinor for general $G$ is deduced using parity arguments. We then
  present the general solution. Plots of energy against
fermion-Skyrmion coupling constant are also given. In section
\ref{deg}, we discuss the degeneracy of energy eigenvalues.
In section \ref{Dirac sea}, we address the problem of calculating the
energy of the Dirac sea using zeta function regularization.  We end
with a conclusion.

\section{The Dirac equation on $S^3$}
\label{Dirac}
Following \cite{Krusch:2003xh}, we now recall the Dirac equation on
a 3-sphere of radius $L=1$.
Consider the stereographic projection from the north pole
$N$ to the plane through the equator.
Let $S^3$ be embedded in
$\mathbb R^4$ with coordinates $(x_1,x_2,x_3,w)$. As a result of
projecting from $N$ onto the equatorial $\mathbb R^3$
plane, points of $S^3$ can be labelled with coordinates $X_i$. The
chart is defined everywhere apart from the projection point, $N$.
The coordinates $X_i$ can be written in terms of $\mathbb R^4$
coordinates as
\begin{equation}
X_i=\frac{x_i}{1-w}.
\end{equation}
We define $R^2=X_{1}^{2}+X_{2}^{2}+X_{3}^{2}$. Then the metric can
be written as
\begin{equation}
g_{{\mathbb R}\times S^3} = {\rm diag}\ \left(1, -\frac{4}{(1+R^2)^2},
-\frac{4}{(1+R^2)^2}, -\frac{4}{(1+R^2)^2} \right).
\end{equation}
We now choose the non-coordinate basis
\begin{equation}
{\hat e}_{\alpha}={e_{\alpha}}^\mu\partial_{X_{\mu}}.
\end{equation}
It is convenient to choose diagonal vierbeins ${e_{\alpha}}^{\mu}$,
such that
\begin{equation}
{e_0}^0=1, \quad {e_i}^i=-\frac{1+R^2}{2},
\end{equation}
where all other components vanish.
With our choice of vierbeins, we can calculate the matrix valued
connection 1-form $\omega_{\alpha\beta}$. The 1-form
$\omega_{\alpha\beta}$ satisfies the metric compatibility condition
$\omega_{\alpha\beta}=-\omega_{\beta\alpha}$, and the torsion-free
condition
\begin{equation}
d{\hat \theta}^{\alpha}+{\omega^{\alpha}}_{\beta}\wedge{\hat
  \theta}^{\beta}=0,
\end{equation}
where ${\hat \theta}^{\alpha}={e^{\alpha}}_{\mu}dX^{\mu}$ is the dual
basis of ${\hat e}_{\alpha}$.
After a short calculation we find
\begin{equation}
\omega^{\alpha\beta} = \left\{ \begin{array}{cl}
  \frac{2}{1+R^2}(X^{\alpha}dX^{\beta}-X^{\beta}dX^{\alpha}) &
  \alpha,\beta = {1,2,3,} \\ 0 & \rm{otherwise}. \end{array} \right.
\end{equation}
The spin connection $\Omega_{\mu}$ can now be expressed as
\begin{equation}
\Omega_{\mu}dX^{\mu}=-\tfrac{i}{2}\omega^{\alpha\beta}\Sigma_{\alpha\beta},
\end{equation}
where
$\Sigma_{\alpha\beta}=\tfrac{i}{4}[\gamma_{\alpha},\gamma_{\beta}]$ and
the components of the commutator are the  standard gamma-matrices,
satisfying $\{\gamma_{\alpha},\gamma_{\beta}\}=2\eta_{\alpha\beta}$.
We work with the following representation of gamma-matrices
\begin{equation}
\label{gamma_matrices}
\gamma^{0} = \left( \begin{array}{cc} 1_{2} & 0\\0 &
  -1_{2}\end{array}\right), \quad  \gamma^{i} = \left(
\begin{array}{cc} 0 & \sigma_{i}\\-\sigma_{i} & 0\end{array}\right),
  \quad \gamma^{5} = \left( \begin{array}{cc} 0 & 1_{2}\\1_{2} &
    0\end{array}\right),
\end{equation}
because we will be working with parity eigenfunctions.
Here $\sigma_{i}$ denotes the set of three Pauli matrices, defined by
\begin{equation}
\sigma_{1} = \left( \begin{array}{cc} 0 & 1\\1 & 0\end{array}\right),
  \quad  \sigma_{2} = \left( \begin{array}{cc} 0 & -i\\i &
    0\end{array}\right), \quad \sigma_{3} = \left( \begin{array}{cc} 1
      & 0\\0 & -1\end{array}\right).
\end{equation}
For massless fermions in curved space-time, the Lagrangian is
\begin{equation}
{\cal
  L}_{fermion}=
\bar{\psi}(i\gamma^{\alpha}{e_{\alpha}}^{\kappa}(\partial_{\kappa}
+\Omega_{\kappa}))\psi.
\end{equation}
With our choice of coordinates and vierbeins, we obtain
\begin{equation}
\label{Lfermion}
{\cal L}_{fermion} = \bar
{\psi}(X_{i},t)\left(i\gamma^{0}\partial_{t}
-i\gamma^{i}\left(\frac{1+R^{2}}{2}\partial_{X_{i}}
-X_{i}\right)\right)\psi(X_{i},t).
\end{equation}
In this paper, we investigate fermions coupled to Skyrmions on $S^{3}$.
We consider a background $B=1$ Skyrme field coupled to the
fermions. The full Lagrangian $\cal L$ is the sum of the fermion
Lagrangian ${\cal L}_{fermion}$, the Skyrmion Lagrangian ${\cal
  L}_{Skyrmion}$ and the interaction Lagrangian ${\cal L}_{int}$. We
consider fermions in the background of a static Skyrme field and
neglect the backreaction. Therefore, we no longer discuss the Skyrmion
Lagrangian, and the interested reader is referred to \cite
{Krusch:2000gb}. ${\cal L}_{int}$ is derived in
\cite{GellMann:1960np}, namely
\begin{equation}
\label{Lint}
{\cal L}_{int}=-g\bar{\psi}(\sigma +i\gamma_{5}\bm{\tau \cdot \pi})\psi,
\end{equation}
where $U=\sigma +i\bm{\tau\cdot\pi}$ is a parametrization of the
Skyrme field and $g$ is the coupling constant.
$\psi$ is a spin-isospin spinor. It is convenient to split the spinor
into two $2\times 2$ spin-isospin matrices $\psi_{1}$ and $\psi_{2}$
such that
\begin{equation}
\textstyle{\psi=\left( \begin{array}{c}\psi_{1}\\\psi_{2} \end{array}
  \right)}.
\end{equation}
Since any complex $2\times 2$ matrix can be expressed as a linear
combination of the Pauli matrices and the identity, it is convenient
to choose these four as a basis of $SU(2)$. The spin-isospin matrices
can then be written as
$\psi_{i}=a_{0}^{(i)}1_{2}+ia_{k}^{(i)}\sigma_{k}$. With this notation
spin operators act on $\psi$ by left multiplication,
$\sigma_{k}\psi_{i}$, whereas the isospin matrices act on $\psi$ by
right multiplication,
\begin{eqnarray}
\nonumber
\tau_{k}\psi_{i}&=&\psi_{i}\sigma_{k}^{T} \\
&=&-\psi_{i}\sigma_{2}\sigma_{k}\sigma_{2}.
\end{eqnarray}

In this paper,
we only consider spherically symmetric Skyrmions. The $B=1$ Skyrmion
on $S^3$ is spherically symmetric \cite {Manton:1987xt}, but for $B
> 1$ this is no longer true. Spherically symmetric Skyrme fields are
best expressed in terms of polar coordinates,
\begin{equation}
U=\exp(if(\mu) {\bm e}_{\mu}\cdot{\bm \tau}),
\end{equation}
where $f(\mu)$ is the ``radial'' shape function and ${\bf e}_{\mu}$ is
the unit vector in the $\mu$ direction, see equation (\ref{emu}).
Using (\ref{Lfermion}) and (\ref{Lint}) we can write down the Dirac
equations for fermions coupled to a spherically symmetric background
Skyrmion. We obtain
\begin{equation}
\label{Diraceq}
\left(
i\gamma^{0}\partial_{t}-
i\gamma^{i}\left(\frac{1+R^2}{2}\partial_{X_{i}}-X_{i}\right)
-gU^{\gamma_{5}}
\right)
\psi(X_{i},t)=0,
\end{equation}
where
\begin{equation}
U^{\gamma_5}=\cos{f(\mu)}+i\gamma_{5}{\bm e}_{\mu}\cdot{\bm \tau}\sin{f(\mu)}.
\end{equation}

\subsection{Solutions of the Dirac Equation for $G=0$}
\label{G=0}
The above Dirac equation (\ref{Diraceq}) and the ansatz $\psi(X_i,t) =
{\rm e}^{i Et} \psi(X_i)$ lead us to the time-independent Dirac equation
\begin{equation}
\label{Schreq}
E\psi=\left( \begin{array}{cc} g\cos{f(\mu)} &
  \bm{\sigma}\cdot\bm{p}+ig\bm{ e}_{\mu}\cdot\bm{\tau}
  \sin{f(\mu)}\\
\bm{ \sigma}\cdot\bm{p}-ig\bm{ e}_{\mu}\cdot\bm{\tau}
  \sin{f(\mu)} & -g\cos{f(\mu)}\end{array}\right)\psi,
\end{equation}
where
\begin{equation}
\label{emu}
\textstyle{{\bm e}_{\mu}=\left(
  \begin{array}{c}\sin\theta\cos\phi\\\sin\theta\sin\phi\\\cos\theta
  \end{array} \right), \quad {\bm e}_{\theta}= \left(
  \begin{array}{c}\cos\theta\cos\phi\\\cos\theta\sin\phi\\-\sin\theta
  \end{array} \right), \quad {\bm e}_{\phi}= \left(
  \begin{array}{c}-\sin\phi\\\cos\phi\\0 \end{array} \right)},
\end{equation}
and
\begin{equation}
\textstyle{\bm{ \sigma}\cdot{\bf p}=-i\left(\bm{
    e}_{\mu}\cdot\bm{\sigma}\left(\partial_{\mu}+
\frac{\sin{\mu}}{1-\cos{\mu}}\right)-\frac{1}{\sin{\mu}}{\bm
    e}_{\theta}\cdot{\bm{
    \sigma}\partial_{\theta}-\frac{1}{\sin{\mu}\sin{\theta}}{\bm
    e}_{\phi}}\cdot\bm{\sigma}\partial_{\phi}\right)}.
\end{equation}
The elements of the matrix in (\ref{Schreq}) commute with the total
angular momentum operator ${\bf G} ={\bf{L+S+I}}$ where ${\bf L}$ is
the orbital angular momentum, ${\bf S}=\tfrac{1}{2}{\bm \sigma}$ is the
spin operator and ${\bf I}=\tfrac{1}{2}{\bm \tau}$ is the isospin
operator.

The above equation (\ref{Schreq}) is invariant under parity $\hat{P}$
where
\begin{equation}
\textstyle{\hat{P}\psi(X_{i})=\gamma_{0}\psi(-X_{i}), \quad
  \hat{P}X_{i}\hat{P}^{-1}=-X_{i}}.
\end{equation}
The $G=0$ case is treated in \cite {Krusch:2003xh}. There the ansatz
for $\psi$ gives rise to a system of two first order ODEs, which can
be expressed as a second order ODE. This equation can be solved
analytically for $f(\mu)=0$ and $f(\mu)=\mu$.
In \cite {Krusch:2003xh} the following energy spectrum was derived
for $f(\mu)=0$,
\begin{equation}
\label{Ef=0}
\textstyle{E=\pm\sqrt{g^{2}+\left (N+\tfrac{3}{2}\right )^{2}} \quad
  {\rm for} \quad N=0,1,2,\dots}
\end{equation}
Setting $u=\cos{\mu}$, the eigenfunctions $G_{N}(u)$ were found to be
given by Jacobi polynomials. The shape function $f(\mu)=\mu$ was also
considered in \cite{Krusch:2003xh}. This
leads to a second order Fuchsian equation with four regular singular
points, two at $u=\pm 1$, one at infinity and one depending on $E$ and
$g$.  The equation could still be solved in terms of
polynomials. The following energy spectrum was derived,
\begin{equation}
\label{Ef=mu}
\textstyle{E_{0}=\frac{3}{2}-g, \quad E_{n}^{\pm}=\frac{1}{2}\pm
  \sqrt{n^{2}+2n+(g-1)^{2}} \quad {\rm for} \quad n=1,2,\dots}
\end{equation}
with eigenfunctions
\begin{equation}
G_{n}(u)=\sum_{j=0}^{n}a_{j}(u+1)^j,
\end{equation}
where
\begin{equation}
\label{aj1}
\textstyle{a_j =
  \frac{(-1)^j(E+g-\frac{3}{2})(E-g+\frac{2j+1}{2})}{j!(2j+1)!!}
\prod\limits_{i=1}^{j-1}(E^2-E+2g-g^2+\frac{1}{4}-(i+1)^2)},
\end{equation}
for $j=1,2,\dots$ and $a_0=1.$ Here $(2j+1)!! = 1\cdot 3 \cdot \dots
\cdot (2j+1)$ is the product of odd integers.
Using $E=\tfrac{1}{2}\pm\sqrt{(n+1)^2-2g+g^2}$ in the product in
(\ref{aj1}), $a_j$ can be written as
\begin{equation}
\label{aj2}
\textstyle{a_j = \frac{(-1)^j(E+g-\frac{3}{2})
(E-g+\frac{2j+1}{2})}{j!(2j+1)!!}\prod\limits_{i=1}^{j-1}((n-i)(n+i+2))}.
\end{equation}
Expanding the product in (\ref{aj2}) we obtain
\begin{equation}
\label{aj3}
\textstyle{a_j = (-1)^j\left(\begin{array}{c}  n\\j
  \end{array}\right)\frac{(n+j+1)!}{(n+1)!(2j+1)!!}
\frac{(E+g-\frac{3}{2})(E-g+\frac{2j+1}{2})}{n(n+2)}}.
\end{equation}

\section{Solutions of the Dirac equation for general $G$}
\label{G}
In the following we derive the solution of the Dirac equation
(\ref{Schreq}) for general $G$.
As a starting point, we construct the total angular momentum operator
eigenstates $\left|jm \right >_{1}$ and $\left|jm \right >_{2}$ in terms
of angular momentum and spin states.  They are expressed as
\begin{eqnarray}
&&
\textstyle{\left|jm \right
  >_{1} = \sqrt{\frac{j-m}{2j}}Y_{j-\frac{1}{2},m+\frac{1}{2}}
  \left|\frac{1}{2} -\frac{1}{2} \right>_{S} +
  \sqrt{\frac{j+m}{2j}}Y_{j-\frac{1}{2},m-\frac{1}{2}}
\left|\frac{1}{2} \frac{1}{2}\right>_{S}},\\
&&
\textstyle{\left|jm
  \right>_{2} = \sqrt{\frac{j+m+1}{2j+2}}Y_{j+\frac{1}{2},m+\frac{1}{2}}
\left|\frac{1}{2} -\frac{1}{2}\right>_{S} -
\sqrt{\frac{j-m+1}{2j+2}}Y_{j+\frac{1}{2},m-\frac{1}{2}}\left|\frac{1}{2}
\frac{1}{2}\right>_{S}},
\end{eqnarray}
where $Y_{j,m}$ is a spherical harmonic and $\left|\frac{1}{2}
\pm\frac{1}{2}\right>_{S}$ is a spin state.
For general $G$ we consider four eigenstates, each of which can be
written in terms of $\left|jm\right>_{1}$ and
$\left|jm\right>_{2}$. These are
\begin{eqnarray}
\label{Gstates1}
\textstyle{\left|GM\right>_{a,c}=
\sqrt{\frac{G-M}{2G}}\left|j=G-\frac{1}{2},m=M+\frac{1}{2}\right>_{1,2}
\left|\frac{1}{2}-\frac{1}{2}\right>_{I}}\nonumber\\
\textstyle{+ \sqrt{\frac{G+M}{2G}}\left|j=G-\frac{1}{2},m=M-
\frac{1}{2}\right>_{1,2}\left|\frac{1}{2} \frac{1}{2}\right>_{I}},
\end{eqnarray}
and
\begin{eqnarray}
\label{Gstates2}
\textstyle{\left|GM\right>_{b,d}=\sqrt{\frac{G+M+1}{2G+2}}
\left|j=G+\frac{1}{2},m=M+\frac{1}{2}\right>_{1,2}\left|\frac{1}{2}
  -\frac{1}{2}\right>_{I}}\nonumber\\
\textstyle{-
  \sqrt{\frac{G-M+1}{2G+2}}\left|j=G+\frac{1}{2},m=M-\frac{1}{2}\right>_{1,2}
\left|\frac{1}{2} \frac{1}{2}\right>_{I}},
\end{eqnarray}
where $\left|\frac{1}{2} \pm\frac{1}{2}\right>_{I}$ is an isospin state.
To carry out our analysis, an ansatz in terms of the $G$-eigenstates
must be found. For each row of our ansatz for the spin-isospin spinor
the states must be of the same parity. Under parity
\begin{equation}
Y_{l,m} \rightarrow (-1)^{l}Y_{l,m},
\end{equation}
so that
\begin{equation}
\textstyle{\left|jm \right>_{1,2} \rightarrow (-1)^{j \mp
\frac{1}{2}}\left|jm \right>_{1,2}}.
\end{equation}
It follows that
\begin{equation}
\begin{array}{ll}
\textstyle{\left|GM\right>_{a}\rightarrow
  (-1)^{G-1}\left|GM\right>_{a}}, \quad&
\textstyle{\left|GM\right>_{b}\rightarrow (-1)^{G}\left|GM\right>_{b}}, \\
\textstyle{\left|GM\right>_{c}\rightarrow
  (-1)^{G}\left|GM\right>_{c}},
\quad& \textstyle{\left|GM\right>_{d}\rightarrow
  (-1)^{G+1}\left|GM\right>_{d}}.
\end{array}
\end{equation}
We see that $\left|GM\right>_{b}$ and $\left|GM\right>_{c}$ both have
parity $(-1)^G$ and that $\left|GM\right>_{a}$ and
$\left|GM\right>_{d}$ both have parity $-(-1)^G$. Hence the spinor
\begin{equation}
\psi = \left( \begin{array}{c}
  b(u)\left|GM\right>_{b}+c(u)\left|GM\right>_{c} \\
  d(u)\left|GM\right>_{d}+a(u)\left|GM\right>_{a} \end{array} \right),
\end{equation}
will have overall parity $(-1)^G$. Clearly exchanging the upper and
lower rows will change the parity by a factor $-1$.  A short
calculation shows that this is equivalent to making the transformation
\begin{equation}
\label{gsym}
g \rightarrow -g
\end{equation}
in the resulting equations. For general $G$ and parity $(-1)^G$
we make the ansatz
\begin{equation}
\label{psi}
\psi = \left( \begin{array}{c} \sqrt{1-u}\sqrt{1-u^2}^G
  G_{2}(u)\left|GM\right>_{b}+\sqrt{1+u}\sqrt{1-u^2}^{G-1}G_{3}(u)
\left|GM\right>_{c} \\ i \sqrt{1+u}\sqrt{1-u^2}^G G_{4}(u)
\left|GM\right>_{d}+i \sqrt{1-u}\sqrt{1-u^2}^{G-1} G_{1}(u)
\left|GM\right>_{a} \end{array} \right),
\end{equation}
where $G_1(u)$, $G_2(u)$, $G_3(u)$ and $G_4(u)$ are functions of $u$ to be
found, and the normalization factors are chosen for later convenience.
Substituting this state into (\ref{Schreq}), we obtain a system of four
coupled first order differential equations in $G_1(u)$, $G_2(u)$,
$G_3(u)$ and $G_4(u)$.  We will solve for the case $f(\mu)=\mu$.

We need to know how the operator ${\bm e_{\mu}}\cdot{\bm \sigma}$ acts
on the $G$-eigenstates in (\ref{Gstates1}) and (\ref{Gstates2}).
The results we require are
\begin{eqnarray}
\label{emusigma1}
\textstyle{{\bm e}_{\mu}\cdot{\bm {\sigma}}\left |GM\right
  >_a=-\left|GM\right>_c}, \quad \textstyle{{\bm e}_{\mu}\cdot{\bm
    {\sigma}}\left |GM\right >_c=-\left |GM\right >_a},\\
\label{emusigma2}
\textstyle{{\bm e}_{\mu}\cdot{\bm {\sigma}}\left |GM\right
  >_b=-\left|GM\right>_d}, \quad \textstyle{{\bm e}_{\mu}\cdot{\bm
    {\sigma}}\left|GM\right>_d=-\left|GM\right>_b}.
\end{eqnarray}
These results can be derived by first deducing that
\begin{eqnarray}
\label{emua}
{\bm e}_{\mu}\cdot{\bm {\sigma}}\left |jm\right
>_1&=&-\left|jm\right>_2,\\
\label{emub}
{\bm e}_{\mu}\cdot{\bm {\sigma}}\left |jm\right >_2&=&-\left|jm\right>_1.
\end{eqnarray}
To obtain the relations (\ref{emua}) and (\ref{emub})
the operator ${\bm e}_{\mu}\cdot{\bm
  {\sigma}}$ is expressed in terms of spherical harmonics. Formulae
for products of spherical harmonics are then needed. The required
results can be found in \cite{a and w}, (\ref{emusigma1}) and
  (\ref{emusigma2}) then follow.

We also need to know how the operator $(-{\bm e}_{\theta}\cdot{\bm
  {\sigma}}\partial_{\theta}-\frac{1}{\sin \theta}{\bm
  e}_{\phi}\cdot{\bm {\sigma}}\partial_{\phi})$ acts on the
$G$-eigenstates. The results are
\begin{eqnarray}
\label{2LS1}
&&
\textstyle{(-{\bm e}_{\theta}\cdot{\bm
    {\sigma}}\partial_{\theta}-\frac{1}{\sin \theta}{\bm
    e}_{\phi}\cdot{\bm {\sigma}}\partial_{\phi})\left|GM \right
  >_{a}=-(G-1)\left|GM \right >_{c}},\\
\label{2LS2}
&&
\textstyle{(-{\bm e}_{\theta}\cdot{\bm
    {\sigma}}\partial_{\theta}-\frac{1}{\sin \theta}{\bm
    e}_{\phi}\cdot{\bm {\sigma}}\partial_{\phi})\left|GM \right
    >_{b}=-G\left|GM \right >_{d}},\\
\label{2LS3}
&&
\textstyle{(-{\bm e}_{\theta}\cdot{\bm
    {\sigma}}\partial_{\theta}-\frac{1}{\sin \theta}{\bm
    e}_{\phi}\cdot{\bm {\sigma}}\partial_{\phi})\left|GM \right
    >_{c}=(G+1)\left|GM \right >_{a}},\\
\label{2LS4}
&&
\textstyle{(-{\bm e}_{\theta}\cdot{\bm
    {\sigma}}\partial_{\theta}-\frac{1}{\sin \theta}{\bm
    e}_{\phi}\cdot{\bm {\sigma}}\partial_{\phi})\left|GM \right
    >_{d}=(G+2)\left|GM \right >_{b}}.
\end{eqnarray}
In order to derive equations (\ref{2LS1})-(\ref{2LS4}) we note that
\begin{equation}
2{\bf{L}\cdot\bf{S}}=S_{+}L_{-}+S_{-}L_{+}+2S_{3}L_{3},
\end{equation}
where $S_{+}$ and $S_{-}$ are defined as $S_{+}=S_{1}+iS_{2}$ and
$S_{-}=S_{1}-iS_{2}$ and $(S_{1},S_{2},S_{3})$ are a set of generators
of the Lie algebra of $SU(2)$ and are related to the Pauli matrices
via $S_{i}=\frac{1}{2}\sigma_{i}$. $L_{+}$, $L_{-}$ and $L_{3}$ are
the orbital angular momentum operators. We note the following result
\begin{equation}
\textstyle{(-{\bm e}_{\theta}\cdot{\bm {\sigma}}\partial_{\theta}-\frac{1}{\sin
    \theta}{\bm e}_{\phi}\cdot{\bm {\sigma}}\partial_{\phi})=({\bm
    e}_{\mu}\cdot{\bm {\sigma}})(2{\bf{L}\cdot\bf{S}})},
\end{equation}
which can easily be proved by multiplying out. Then
\begin{equation}
2{\bf{L}\cdot\bf{S}}\left|jm\right >_1=(j-\tfrac{1}{2})\left|jm\right >_1,
\end{equation}
\begin{equation}
2{\bf{L}\cdot\bf{S}}\left|jm\right >_2=-(j+\tfrac{3}{2})\left|jm\right >_2,
\end{equation}
can be proved by considering how $L_{+}$, $L_{-}$ and $L_{3}$ act on
the spherical harmonics. The necessary formulae can be found in
\cite{a and w}. These two equations also follow from
$2{\bf{L}\cdot\bf{S}}={\bf{J}}^{2}-{\bf{L}}^{2}-{\bf{S}}^{2}$.

It can then be seen that
\begin{equation}
(-{\bm e}_{\theta}\cdot{\bm {\sigma}}\partial_{\theta}-\tfrac{1}{\sin
    \theta}{\bm e}_{\phi}\cdot{\bm
    {\sigma}}\partial_{\phi})\left|jm\right
  >_1=-(j-\tfrac{1}{2})\left|jm\right >_2 ,
\end{equation}
\begin{equation}
(-{\bm e}_{\theta}\cdot{\bm {\sigma}}\partial_{\theta}-\tfrac{1}{\sin
    \theta}{\bm e}_{\phi}\cdot{\bm
    {\sigma}}\partial_{\phi})\left|jm\right
  >_2=(j+\tfrac{3}{2})\left|jm\right >_1 .
\end{equation}
This leads to (\ref{2LS1})-(\ref{2LS4}).

The operator ${\bm e_{\mu}}\cdot{\bm \tau}$ acts on the
$G$-eigenstates to give
\begin{eqnarray}
\label{emutau1}
&&
{\bm e_{\mu}}\cdot{\bm
  \tau}\left|GM\right>_{a}=-\frac{2\sqrt{G(G+1)}}{2G+1}
\left|GM\right>_{b}-\frac{1}{2G+1}\left|GM\right>_{c},\\
\label{emutau2}
&&
{\bm e_{\mu}}\cdot{\bm
  \tau}\left|GM\right>_{b}=-\frac{2\sqrt{G(G+1)}}{2G+1}
\left|GM\right>_{a}+\frac{1}{2G+1}\left|GM\right>_{d},\\
\label{emutau3}
&&
{\bm e_{\mu}}\cdot{\bm \tau}\left|GM\right>_{c}=-\frac{1}{2G+1}
\left|GM\right>_{a}-\frac{2\sqrt{G(G+1)}}{2G+1}\left|GM\right>_{d},\\
\label{emutau4}
&&
{\bm e_{\mu}}\cdot{\bm
  \tau}\left|GM\right>_{d}=\frac{1}{2G+1}
\left|GM\right>_{b}-\frac{2\sqrt{G(G+1)}}{2G+1}\left|GM\right>_{c}.
\end{eqnarray}
These equations can be proved by expanding ${\bm e_{\mu}}\cdot{\bm
  \tau}$ then multiplying this, from the right, by each
$G$-eigenstate. The right hand side of each equation is then computed
and matrix components are compared.

Substituting (\ref{psi}) into (\ref{Schreq}) and using the identities
  (\ref{emusigma1}), (\ref{emusigma2}), (\ref{2LS1})-(\ref{2LS4}) and
  (\ref{emutau1})-(\ref{emutau4}) we
  are led to two equations, one in $\left|GM\right>_{a}$ and
  $\left|GM\right>_{d}$ and one in $\left|GM\right>_{b}$ and
  $\left|GM\right>_{c}$. We equate coefficients of the $G$-eigenstates
  to obtain four ODEs in $G_{1}(u),$ $G_{2}(u),$ $G_{3}(u)$ and
  $G_{4}(u)$. These ODEs are shown below,
\begin{footnotesize}
\begin{eqnarray}
\label{dG1du}
 (1-u)\frac{dG_1}{du} & = &
 \left(G+\tfrac{1}{2}-\frac{g(1-u)}{2G+1}\right)G_1+(E-gu)G_3
-\frac{2g\sqrt{G(G+1)}(1-u^2)}{2G+1}G_4,
 \\
\label{dG2du}
 (1-u)\frac{dG_2}{du} & = &
 \left(G+\tfrac{3}{2}-\frac{g(1-u)}{2G+1}\right)G_2-(E+gu)G_4
+\frac{2g\sqrt{G(G+1)}}{2G+1}G_3,
 \\
\label{dG3du}
  (1+u)\frac{dG_3}{du} & = &
 -\left(G+\tfrac{1}{2}-\frac{g(1+u)}{2G+1}\right)G_3-(E+gu)G_1
+\frac{2g\sqrt{G(G+1)}(1-u^2)}{2G+1}G_2,\\
\label{dG4du}
  (1+u)\frac{dG_4}{du} & = &
 -\left(G+\tfrac{3}{2}-\frac{g(1+u)}{2G+1}\right)G_4+(E-gu)G_2
-\frac{2g\sqrt{G(G+1)}}{2G+1}G_1.\
\end{eqnarray}
\end{footnotesize}

\noindent
These equation give the solutions for states with parity $(-1)^G$. Due
to the symmetry (\ref{gsym}) states with parity $(-1)^{G+1}$ are
obtained by replacing $g$ by $-g$ in the equations above.

\subsection{The energy spectrum}
In this section, we derive the energy spectrum of the
time-independent Dirac equation (\ref{Schreq}) and the corresponding
eigenfunctions. We discuss a useful symmetry of our system of
equations (\ref{dG1du})-(\ref{dG4du}) and also comment on associated
second order and fourth order equations. In order to derive the
spectrum, we use the theory of Fuchsian differential equations,
and in particular, regular singular points and their exponents, see
\cite{whittaker}. Finally, we present the explicit solution.

Under the transformation
\begin{equation}
\label{sym}
\left(G_1(u), G_2(u), G_3(u), G_4(u)\right) \mapsto
\left(-G_3(-u),-G_4(-u),-G_1(-u),-G_2(-u)\right)
\end{equation}
followed by $u \mapsto -u,$ (\ref{dG2du}) is mapped
into (\ref{dG4du}) and (\ref{dG1du}) into (\ref{dG3du}), and vice
versa. Hence the system of equations (\ref{dG1du})-(\ref{dG4du})
remains invariant.
Eliminating $G_1(u)$ and $G_3(u)$ from the system
(\ref{dG1du})-(\ref{dG4du}) results in a system
of two second order equations, which again map into each other via the
symmetry (\ref{sym}). These second order equations prove to be useful
for deriving an ansatz for $G_2(u)$ and $G_4(u).$
Finally, we can also derive two fourth order
ODEs, by eliminating $G_4(u)$ and $G_2(u)$, respectively, again related
via (\ref{sym}). Hence, if we have solutions for
$G_2(u)$ we can find solutions for $G_4(u)$. We can then use our system of
first order ODEs to find solutions for $G_1(u)$ and $G_3(u)$.

In order to solve our equations we first have to derive the energy
eigenvalues.
Our system of equations (\ref{dG1du})-(\ref{dG4du}) has
regular singular points at $u=\pm 1$ and an
irregular singular point at infinity. For both regular singular points
the exponents are
\begin{equation}
(0, 0, -G-\tfrac{1}{2}, -G-\tfrac{3}{2}).
\end{equation}
We require our solutions to be regular over the whole 3-sphere and, in
particular, at the north and south poles, $u=\pm 1$.  The solutions
corresponding to the exponents $-G-\frac{1}{2}$ and $-G-\frac{3}{2}$
contain poles, so we can exclude these solutions. The fact that there
are two exponents taking values of zero means that corresponding to
each regular singular point is a solution with logarithmic terms and
the solution is therefore singular. As a result, we can also
exclude these solutions. The regular solution can therefore be
expanded as a power series in $1+u$ around the south pole, and also a
power series in $1-u$ around the north pole. These two expansions only
agree for certain values of the energy $E$. It turns out that these
energy eigenvalues can be calculated from the exponents at infinity of
the fourth order ODE in $G_2(u)$, mentioned above,
which arises by eliminating $G_1(u),$
$G_3(u),$ and $G_4(u)$ from our system of equations
(\ref{dG1du})-(\ref{dG4du}).
This equation is of Fuchsian type and has three regular singular points
at $u=\pm 1$ and infinity.

The solutions of Fuchsian differential equations can only have
singularities at their singular points. According to our discussion
above, we are interested in the solutions of (\ref{Schreq}) which are
non-singular. Therefore, $G_2(u)$ has to be regular at $u=\pm 1$, and
hence on the entire complex plane. So, $G_2(u)$ is an analytic
function, in fact an integral function, on the complex plane. As it is
the solution of a Fuchsian differential equation, it can only have
poles at infinity, and it follows that $G_2(u)$ is a polynomial.

The exponents corresponding to $u=\infty$ can be found by setting
$u=\frac{1}{z}$ then considering $z\rightarrow 0$. We obtain the exponents
\begin{eqnarray}
\label{rhos}
\rho_s &=& 1+G\pm\tfrac{1}{2}\sqrt{1+4E^{2}+4E-4g^{2}},\\
\label{rhoa}
\rho_a &=& 1+G\pm\tfrac{1}{2}\sqrt{1+4E^{2}-4E+8g-4g^{2}}.
\end{eqnarray}

As argued above $G_2(u)$ is a polynomial. Let its degree be denoted by $k.$
Then the exponents at $u=\infty$ can be equated with $-k$.
From the exponent $\rho_s$ we obtain the following energy eigenvalues
\begin{equation}
E_{sym}^{\pm}=-\tfrac{1}{2}\pm\sqrt{(k+G+1)^2+g^2}
\quad \rm{for~}G=1,2,\dots,~ k=0,1,\dots
\end{equation}
$E_{sym}^{\pm}$ is a novel feature which arises for $G> 0$ only. Note
that this energy is invariant under $g \mapsto -g$.
From $\rho_a$ in (\ref{rhoa}) we obtain another family of energy
eigenvalues, namely,
\begin{equation}
E_{asym}^{\pm}=\tfrac{1}{2}\pm \sqrt{(k+G+1)^2-2g+g^2} \quad
\rm{for~} G=0, 1,\dots,~ k=0, 1,\dots,
\end{equation}
where $G$ and $k$ are not both zero.
This energy spectrum has already been obtained in \cite{Krusch:2003xh}
for the case $G=0.$  A slight subtlety occurs for $k=0$ and $G=0$. In
this case, only
\begin{equation}
\label{E0}
E_{0}=\tfrac{3}{2}-g
\end{equation}
leads to a regular solution. The energy level (\ref{E0}) is rather
special as it crosses from the positive spectrum to the
negative spectrum as the coupling constant $g$ is varied, also see
\cite{Krusch:2003xh} for further details.

Now that we have derived the energy spectrum, we can solve the
system (\ref{dG1du})-(\ref{dG4du}) by  first considering the
fourth order ODE in $G_2(u)$. We make the ansatz that  $G_2(u)$ is a
polynomial in $1+u$ and insert this into our ODE to find the
polynomial coefficients. The symmetry (\ref{sym}) and the system of
second order equations for $G_2(u)$ and $G_4(u)$
leads to a related expression for $G_4(u)$.
Then the solution corresponding to $E^{\pm}_{asym}$ is found to be
\begin{equation}
G_2(u)=\sum_{j=0}^{k}a_{j}(1+u)^{j} \quad {\rm and} \quad
G_4(u)=(-1)^{k}\sum_{j=0}^{k}a_{j}(1-u)^{j}.
\end{equation}
The general expression for $a_j$ is
\begin{equation}
a_{j}=(-1)^j\left(\begin{array}{c}  k\\j
\end{array}\right)
\tfrac{(2G+k+j+1)!(2G+1)!!}{(2G+k+1)!(2G+2j+1)!!}
\tfrac{(E+g-\tfrac{3}{2}-G)(E-g+\tfrac{2j+1}{2}+G)}{2G(k+g)+k(k+2)}.
\end{equation}
If we set $G=0$, and hence $k=n$, the above formula leads us to (\ref{aj3})
which is equivalent to the result from \cite{Krusch:2003xh}.

For $E^{\pm}_{sym}$ we find
\begin{equation}
G_2(u)=\sum_{j=0}^{k}a_{j}(1+u)^{j} \quad {\rm and} \quad
G_4(u)=(-1)^{k+1}\sum_{j=0}^{k}a_{j}(1-u)^{j}.
\end{equation}
The general expression for $a_j$ is now
\begin{equation}
a_{j}=(-1)^{j+1}\left(\begin{array}{c}  k\\j
\end{array}\right)\tfrac{(2G+k+j+1)!(2G+1)!!}{(2G+k+1)!(2G+2j+1)!!}
\tfrac{(E+g+\tfrac{3}{2}+G)(E-g-\tfrac{2j+1}{2}-G)}{2(G+1)(g-k)-k^2}.
\end{equation}
We then use equations (\ref{dG4du}) and (\ref{dG2du}) to obtain $G_{1}(u)$
and $G_{3}(u),$ respectively, and it is easy to see that $G_1(u)$ and
$G_3(u)$ are polynomials of order $k+1.$

We can carry out a consistency check on our solutions by setting $g=0$
in equations (\ref{dG1du})-(\ref{dG4du})
and manipulating the equations to obtain two uncoupled
second order ODEs. These are both Jacobi equations and have polynomial
solutions which can be expressed in terms of
hypergeometric functions (see \cite{handbook}). For $g=0$, our solutions
are the same.

\subsection{Degeneracy of the energy spectrum}
\label{deg}

\begin{figure}[!htb]
\begin{center}
\begin{includegraphics}[width=12cm]{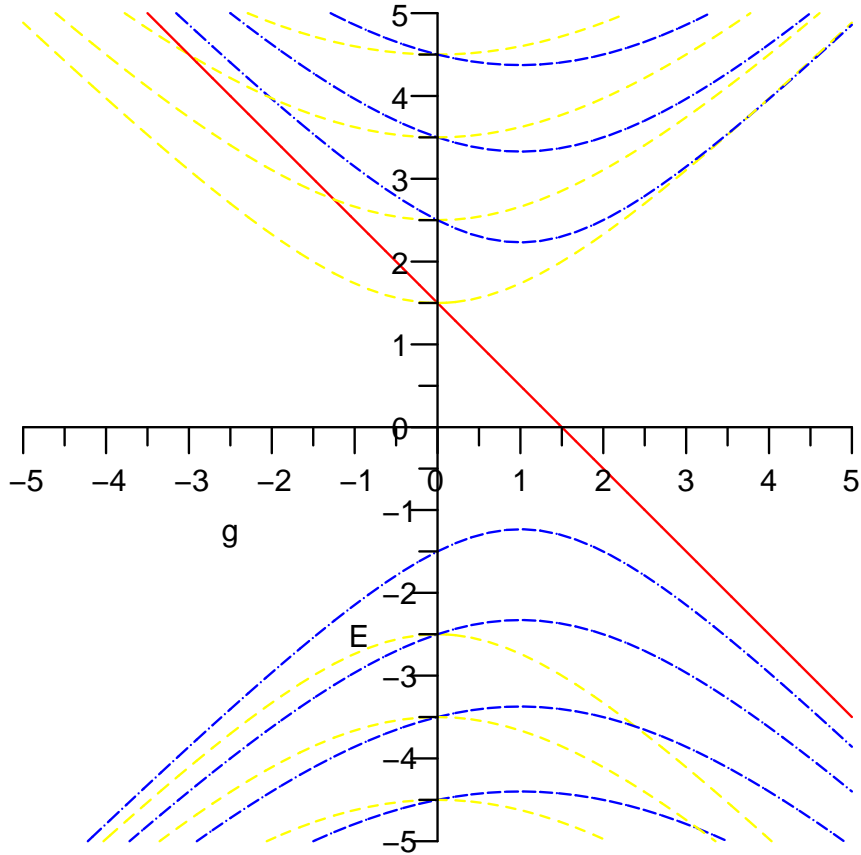}\end{includegraphics}
\caption{The energy $E$ as a function of the coupling constant $g$ for
  $E_0$ (solid red curve), $E_{asym}^{\pm}(n)$ (dotted
  blue curves) and $E_{sym}^{\pm}(n)$ (dashed yellow curves).
  \label{energy}}
\end{center}
\end{figure}

In order to discuss the degeneracy of the energy spectrum
it is convenient to introduce $n=k+G$, where the integer
$n$ is analogous to the
principal quantum number arising in the quantum mechanics of the
hydrogen atom. Then the energy spectrum for states of parity $(-1)^G$
is given by the two families
\begin{eqnarray}
E_{sym}^{\pm}(n)&=&-\tfrac{1}{2}\pm\sqrt{(n+1)^2+g^2}
\quad \rm{for~}n=1,2,\dots, \\
E_{asym}^{\pm}(n)&=&\tfrac{1}{2}\pm \sqrt{(n+1)^2-2g+g^2} \quad
\rm{for~} n= 1,2\dots,
\end{eqnarray}
and the special energy level (\ref{E0}),
\begin{equation*}
E_{0}=\tfrac{3}{2}-g.
\end{equation*}
Figure \ref{energy} shows the energy spectra for different values of
$n$. There are two different ways of reading figure \ref{energy}. The
obvious interpretation is the energy spectrum of states of parity
$(-1)^G$ as a function of the coupling constant $g \in {\mathbb R}.$
For the second interpretation and in the following,
we restrict our attention to $g \ge
0$. Then, the negative values of $g$ corresponds to states with parity
$(-1)^{G+1}$ due to symmetry (\ref{gsym}), while positive values of
$g$ again correspond to states of parity $(-1)^G.$ The latter interpretation
is very useful for discussing the degeneracy of the spectrum.

The energy level (\ref{E0}) only exists for $G=0$. It
gives rise to a positive parity state
with energy $\tfrac{3}{2} - g$ and a negative parity state with
energy $\frac{3}{2} + g.$ Since the degeneracy of a state with grand spin
$G$ is $2G+1$ these two states are non-degenerate for $g>0,$
and ``parity doubling'' occurs for $g=0$, \cite{Balachandran:1998zq}.

For $E^{\pm}_{asym}(n),$ positive and negative parity states will in
general have different energy eigenvalues for a given value of the
coupling constant $g$. Recall that $n=k+G$, hence  $G$ can vary from
$0$ to $n.$ Therefore, the degeneracy is
\begin{equation}
D(E^{\pm}_{asym}(n))=\sum_{G=0}^{n}(2G+1)=(n+1)^2.
\end{equation}

For $E^{\pm}_{sym}(n),$ positive and negative parity states have the
same energy for a given value of the coupling constant $g.$ These
states only exist for $G>0$,  hence $G$ varies from $1$ to $n$.
Therefore, the degeneracy is
\begin{equation}
D(E^{\pm}_{sym}(n))=2\sum_{G=1}^n(2G+1)=2 n(n+2),
\end{equation}
and the extra factor of $2$ is due to parity.

We now consider the case of zero coupling ($g=0$) which is equivalent
to massless fermions on $S^3$. In this case there will clearly always be
invariance under $g \rightarrow -g$, so
\begin{small}
\begin{equation}
D_{g=0}\left(E_{0}\right)= 2,\quad
D_{g=0}\left(E_{asym}^{\pm}(n)\right) = 2\left(n+1\right)^{2},\quad
D_{g=0}\left(E_{sym}^{\pm}({\tilde n})\right) =
2{\tilde n}\left({\tilde n}+2\right).
\end{equation}
\end{small}

\noindent
At $E=\tfrac{3}{2}$ the energy level $E_0$ and $E_{sym}^+(1)$ are
degenerate, hence the degeneracy
is
\begin{equation}
D_{g=0}(E=\tfrac{3}{2})=8.
\end{equation}
The energy eigenvalue
$E=-\tfrac{3}{2}$ only occurs for $E_{asym}^-(1)$, hence
the degeneracy is again
\begin{equation}
D_{g=0}(E=-\tfrac{3}{2})=8.
\end{equation}

The energy eigenvalue $E = N + \tfrac{3}{2},$ $N = 1$, $2$, \dots,
is attained by $E_{asym}^+(N),$ and $E_{sym}^+(N+1),$ hence
\begin{equation}
\label{deg0a}
D_{g=0}\left(E=N+\tfrac{3}{2}\right)=2(N+1)(N+3)+2(N+1)^2 = 4(N+1)(N+2).
\end{equation}
Similarly, $E = -N - \tfrac{3}{2}$ is attained by $E_{asym}^-(N+1)$ and
$E_{sym}^-(N),$ hence
\begin{equation}
\label{deg0b}
D_{g=0}\left(E=-N-\tfrac{3}{2}\right)
=2(N+2)^2+2N(N+2)=4(N+1)(N+2).
\end{equation}
After considering the factor of $2$ due to isospin
and another factor of $2$ due to parity, equations (\ref{deg0a}) and
(\ref{deg0b}) are consistent with the results in
\cite{Camporesi:1995fb}.

So far, we have only considered generic degeneracies and the case
$g=0.$ This energy
spectrum is rather special in that we can also calculate all the
accidental degeneracies for $g>0$. These degeneracies all occur for
rational values of $g.$ For example, the negative parity state with
energy $\tfrac{3}{2} +g$ is only degenerate with the states
with $E_{sym}^+(n)$ for
\begin{equation}
g = \tfrac{1}{4} (n-1)(n+3)
\end{equation}
and with the $(-1)^{G+1}$ parity states with $E_{asym}^-(n)$ (changing
$g$ to $-g$) for
\begin{equation}
g = \tfrac{1}{4} n(n+2).
\end{equation}
The positive parity state with energy $\tfrac{3}{2}-g$ is always
non-degenerate for $g>0.$
Similarly, $(-1)^G$ parity states of energy $E_{asym}^\pm(n)$ are
degenerate with $(-1)^{G+1}$ parity states of energy
$E_{asym}^\pm({\tilde n})$ for
\begin{equation}
g = \tfrac{1}{4} n(n+2) - \tfrac{1}{4} {\tilde n}({\tilde n}+2),
\end{equation}
which is positive for $n>{\tilde n}$.
Finally, states with energy $E_{asym}^+(n)$ and states with energy
$E_{sym}^+({\tilde n})$ are degenerate for
\begin{equation}
g=\frac{4(\tilde n +1)^2-\left(1+(\tilde n +1)^2 -(n+1)^2
\right)^2}{4\left(1+(\tilde n+1)^2
-(n+1)^2\right)},
\end{equation}
and a similar equation holds for $E_{asym}^-(n)$ and
$E_{sym}^-({\tilde n}).$

\section{The Dirac Sea}
\label{Dirac sea}
In this section, we briefly comment on the zeta function regularization
\cite{Elizalde:1994gf}. In order to calculate the energy of the Dirac sea,
\begin{equation}
\label{naive}
E_{{\rm Dirac}} = \sum\limits_{N=0}^\infty D(N) E(N),
\end{equation}
where $E(N)$ is the $N$th negative energy eigenvalue, see
e.g. \cite{Rajaraman:1982is}, and $D(N)$ is its
degeneracy,  we define the zeta function
\begin{equation}
\label{zeta}
\zeta(s) = \sum\limits_{N=0}^\infty D(N) E(N)^{-s}.
\end{equation}
The expression (\ref{naive}) is clearly divergent. However, the expression
(\ref{zeta}) is convergent for large enough $s$.  The Dirac sea energy
(\ref{naive}) is then defined by the analytic continuation of
(\ref{zeta}) to $s=-1$.
For example for $g=0$ we have
\begin{equation}
E_{g=0} = -4\sum\limits_{N=0}^\infty
(N+1)(N+2)(N+\tfrac{3}{2})^{-s}\Big|_{s=-1}.
\end{equation}
Hence, the relevant zeta function is
\begin{equation}
\label{zetag=0}
\zeta_{g=0}(s) = -4\sum\limits_{N=0}^\infty
((N+\tfrac{3}{2})^2-\tfrac{1}{4})(N+\tfrac{3}{2})^{-s},
\end{equation}
which can be rewritten as
\begin{equation}
\zeta_{g=0}(s) = -4 \zeta_H(s-2,\tfrac{3}{2})
+ \zeta_H(s,\tfrac{3}{2}),
\end{equation}
where $\zeta_H(s,a)$ is the Hurwitz zeta function defined as
\begin{equation}
\label{zetaH}
\zeta_H(s,a) = \sum\limits_{n=0}^\infty (n+a)^{-s}.
\end{equation}
Evaluating $\zeta_{g=0}(s)$ at $s=-1$ we obtain
\begin{equation}
\label{Eg=0}
E_{g=0} = \frac{17}{240}.
\end{equation}

For massive fermions on $S^3$ corresponding to the case $f(\mu)=0$ in
Section \ref{G=0}, the energy is given by (\ref{Ef=0}) and the degeneracy
is
\begin{equation}
D(N) = 4(N+1)(N+2).
\end{equation}
Hence the Dirac sea energy is given by
\begin{equation}
\label{Ezetaf=0}
E_{f(\mu)=0} = -4 \sum\limits_{N=0}^\infty (N+1)(N+2)
\left(\left(N+\tfrac{3}{2}\right)^2 + g^2
\right)^{-s}\Big|_{s=-\tfrac{1}{2}}.
\end{equation}
Zeta functions of generalized Epstein-Hurwitz type are of the form
\begin{equation}
\label{Fsab}
F\left(s;a,b^2\right)
= \sum\limits_{n=0}^\infty \left((n+a)^2 + b^2 \right)^{-s}.
\end{equation}
Asymptotic expansions for (\ref{Fsab}) are discussed in
\cite{Elizalde:1994yz, Elizalde:1994gf}.
Here, we are concerned with a generalization of (\ref{Fsab}),
namely,
\begin{equation}
\label{Fmsab}
F^{(m)}\left(s;a,b^2 \right)
= \sum\limits_{n=0}^\infty (n+a)^m \left((n+a)^2 + b^2 \right)^{-s},
\end{equation}
where we assume that $a>0$ and $b \ge 0$.
We follow \cite{Elizalde:1994gf} to derive a formula for
$F^{(m)}(s=-\tfrac{1}{2};a,b^2).$ We first perform a binomial
expansion which is valid for $b < a$ and rewrite (\ref{Fmsab})
as a contour integral
\begin{eqnarray}
F^{(m)}\left(s;a,b^2 \right)
&=& \sum\limits_{n=0}^\infty
\sum\limits_{k=0}^\infty
(-1)^k \frac{\Gamma(s+k)}{\Gamma(k+1) \Gamma(s)}
b^{2k} (n+a)^{-2s-2k+m},
\\
&=& \sum\limits_{n=0}^\infty
\frac{1}{2 \pi i} \int\limits_{C}
\frac{\Gamma(s+z) b^{2z} (n+a)^{-2s-2z+m}} {\Gamma(z+1) \Gamma(s)}
\frac{\pi}{\sin(\pi z)} {\rm d}z.
\end{eqnarray}
Recall that
\begin{equation}
\frac{\pi}{\sin(\pi z)} = \frac{(-1)^k}{z-k} + O(z-k)\quad {\rm
  for}~k\in{\mathbb Z}.
\end{equation}
The contour $C$ encloses
all the non-negative poles of $1/\sin(\pi z)$
with anti-clockwise orientation and can be split into a part
$$
\int\limits_{-z_0+i\infty}^{-z_0-i\infty},
$$
where $0 < z_0 < \tfrac{1}{2},$
and a semi-circle at infinity. The latter does not contribute to the
integral. Now, we can move the sum over $n$ under the integral and use
the definition of the Hurwitz zeta function (\ref{zetaH}) to obtain
\begin{equation}
\label{Fint}
F^{(m)}\left(s;a,b^2 \right)
=\frac{1}{2i} \int \limits_{-z_0+i\infty}^{-z_0-i\infty}
\frac{\Gamma(s+z) \zeta_H(2s+2z-m,a) b^{2z}} {\Gamma(z+1) \Gamma(s)
  \sin(\pi z)} {\rm d}z.
\end{equation}
This can be evaluated by closing the contour again, and using Cauchy's
theorem. This time the contribution of the integral over the
semi-circle at infinity is non-zero. However, it was shown in
\cite{Elizalde:1994yz} that the contribution is very small, so we
neglect it in the following.

From now on, we focus on the physically relevant value $s=-\tfrac{1}{2}$.
The integral (\ref{Fint}) has poles at $z \in {\mathbb Z}$
due to $1/\sin (\pi z)$. Only the non-negative poles contribute,
because of the contour. The gamma function $\Gamma(z -\tfrac{1}{2})$
has poles at $z - \tfrac{1}{2} = 0, -1, -2, \dots$
Only the pole at $z = \tfrac{1}{2}$ lies inside the contour. Finally,
there is a contribution from the pole of the Hurwitz zeta function at
$2z-1-m=1.$ All the poles are simple unless the pole of $\zeta_H$ at
$z=1+\tfrac{m}{2}$ is a non-negative integer. Hence, the integral in
(\ref{Fint}) becomes
\begin{small}
\begin{equation}
F^{(m)}\left(-\tfrac{1}{2};a,b^2 \right) \approx
{\rm Res}_{z=\tfrac{1}{2}} + {\rm Res}_{z=1+\tfrac{m}{2}}
+ \sum\limits_{\substack{k=0\\ k \neq 1 + \tfrac{m}{2}}}^{\infty}
(-1)^k \tfrac{\Gamma\left(k-\tfrac{1}{2}\right)}{k!
  \Gamma\left(-\tfrac{1}{2}\right)} 
\zeta_H(2k-1-m,a) b^{2k},
\end{equation}
\end{small}

\noindent
where the sum arises from the simple poles of $1/\sin(\pi z).$
Note that
\begin{equation}
\Gamma(\epsilon) = \frac{1}{\epsilon} -\gamma + O(\epsilon)
\end{equation}
and
\begin{equation}
\zeta_H(-k,a) = -\frac{B_{k+1}(a)}{k+1},\quad {\rm for}~k\in {\mathbb N},
\end{equation}
where $B_m(a)$ are the Bernoulli polynomials and $\gamma$ is the
Euler-Mascheroni constant. Hence, the residue at $z=\tfrac{1}{2}$ gives
\begin{equation}
{\rm Res}_{z=\tfrac{1}{2}} = \frac{B_{m+1}(a)}{m+1}b.
\end{equation}
Finally, for the residue
at $z= 1 + \tfrac{m}{2}$, we note that
\begin{equation}
\zeta_H(1+\epsilon,a) = \frac{1}{\epsilon} -\Psi(a) +O(\epsilon),
\end{equation}
where $\Psi(a) = \frac{d}{da} \ln \Gamma(a)$ is the digamma function,
see \cite[p. 271]{whittaker}.
The behaviour depends on whether $m$ is even or
odd. For odd $m$ this is just another simple pole, and we obtain
\begin{equation}
{\rm Res}_{z=1+\tfrac{m}{2}} =
(-1)^{\tfrac{m-1}{2}}
\frac{\sqrt{\pi}\Gamma(\tfrac{m+1}{2})}
{4\Gamma(2+\tfrac{m}{2})} b^{2+m}.
\end{equation}
However, for even $m$ there is a double pole, and we have to use
\begin{equation}
{\rm Res}_{z=1+\tfrac{m}{2}} =
\lim\limits_{z \to 1+\tfrac{m}{2}}
\frac{d}{dz}
\left( \left(z-1-\tfrac{m}{2}\right)^2
\frac{\pi \Gamma(s+z)
  \zeta_H(2z-1-m,a)\ b^{2z}} {\Gamma(z+1) \Gamma(-\tfrac{1}{2})
  \sin(\pi z)}\right)
\end{equation}
to obtain
\begin{eqnarray}
\lefteqn{{\rm Res}_{z=1+\tfrac{m}{2}} =
(-1)^{\tfrac{m}{2}} \frac{b^{2+m}\Gamma(\tfrac{m+1}{2})
}
{4m(2+m)\sqrt{\pi}\Gamma(2+\tfrac{m}{2})}}\\
\nonumber
&& *\left(\left(\Psi(\tfrac{m+1}{2})-\Psi(\tfrac{m}{2})- 2 \Psi(a)
+2\ln(b)\right)m(m+2)- 4(1+m)\right).
\end{eqnarray}

\begin{figure}[!htb]
\begin{center}
\begin{includegraphics}[width=12cm]{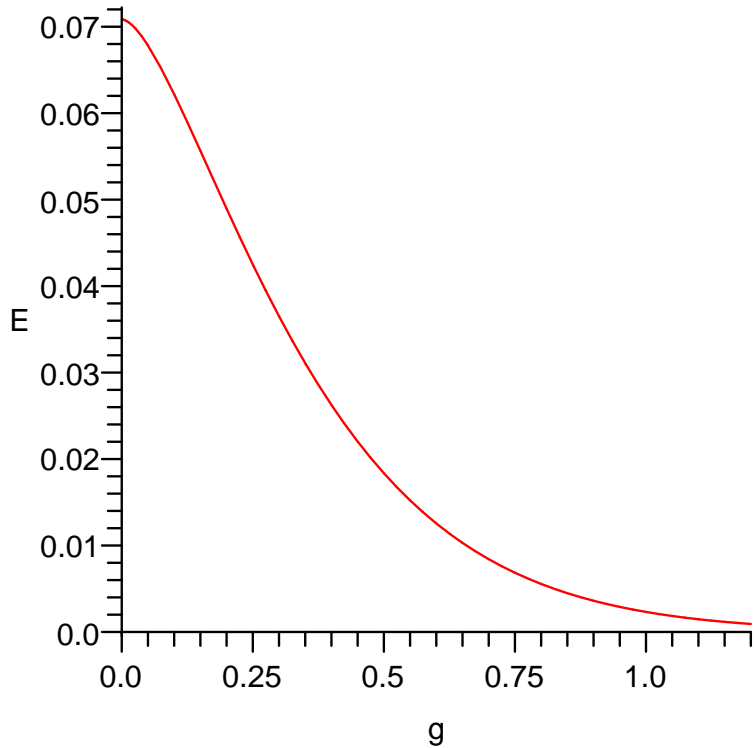}\end{includegraphics}
\caption{The Dirac sea energy $E_{f=0}$ as a function of the coupling
  constant $g$.\label{DiracSea}}
\end{center}
\end{figure}

We now use the same trick as in (\ref{zetag=0}) to rewrite the
regularized energy in (\ref{Ezetaf=0}) as
\begin{equation}
E_{f(\mu)=0} = -4 F^{(2)}(-\tfrac{1}{2};\tfrac{3}{2},g^2)
+ F^{(0)}(-\tfrac{1}{2};\tfrac{3}{2},g^2).
\end{equation}
The regularized energy is plotted in figure \ref{DiracSea}. As a
consistency check it can be shown that $E_{f(\mu)=0}(g=0) = \frac{17}{240}$
as calculated in (\ref{Eg=0}). It would be interesting to compare these
results to other regularization methods.

Finally, we address the problem of calculation the Dirac sea energy
for fermions coupled to a $B=1$ background Skyrmion. In this case the
Dirac Sea energy is given by
\begin{footnotesize}
\begin{eqnarray}
\label{Ezetaf=mu}
\lefteqn{ E_{f(\mu)=\mu}  =  -\sum\limits_{n=1}^\infty
  \left(2n(n+2)\left(\tfrac{1}{2}+\sqrt{\left(n+1\right)^2+g^2}\right)^{-s}
  \right.}  \\
\nonumber
&&
\left.
+ (n+1)^2 \left(-\tfrac{1}{2}+\sqrt{(n+1)^2+2g+g^2}\right)^{-s}
+ (n+1)^2 \left(-\tfrac{1}{2}+\sqrt{(n+1)^2-2g+g^2}\right)^{-s}
\right)\Big|_{s=-1},
\end{eqnarray}
\end{footnotesize}

\noindent
which is the sum of $E_{sym}$ and $E_{asym}$ for both parities. The
energy of the ``zero mode'' $E_0 = \tfrac{3}{2} -g$ also needs to be
taken into account, and we expect a similar picture as in
\cite{Ripka:1985am}.

Unfortunately, this is a much more complicated
situation, and zeta functions of this type have not
been discussed in the literature, to our knowledge. As a starting
point, we could again perform a binomial expansion. We can then
rewrite the energy $E_{f(\mu)=\mu}$ as an infinite sum of zeta
functions $F^{(m)}(-\tfrac{1}{2};a;b^2)$. Unfortunately, the last term
in (\ref{Ezetaf=mu}) leads to $b^2 = -2g+g^2$ which is negative for small
$g$, and our formula no longer converges. It would be interesting to
derive alternative expressions for these types of zeta function.

\section{Conclusion}
\label{Conclusion}
In this paper we consider the Dirac equation for fermions on $S^{3}$
chirally coupled to a spherically-symmetric background
Skyrmion with topological charge one.
The time-independent Dirac equation commutes with the grand
spin and parity, and these symmetries allow us to reduce the Dirac
equation to a system of four linear ODEs.
Making use of the theory
of Fuchsian differential equations, we derive the
complete energy spectrum and the corresponding eigenfunctions which are
given by polynomials. There is a positive parity state with energy
$$
E_0 = \tfrac{3}{2}-g
$$
and a negative parity state with energy
$$
E_0 = \tfrac{3}{2}+g.
$$
Both states are generally non-degenerate.
The energies
$$
E_{asym}^{\pm}(n) = \tfrac{1}{2}\pm \sqrt{(n+1)^2-2g+g^2}
$$
and
$$
E_{asym}^{\pm}(n) = \tfrac{1}{2}\pm \sqrt{(n+1)^2+2g+g^2}
$$
all have degeneracy $(n+1)^2$, and correspond to states with parity
$(-1)^G$ and $(-1)^{G+1},$ respectively. For $G=0,$ these energies
were found in \cite{Krusch:2003xh}. Finally, the energies
$$
E_{sym}^{\pm}(n) = -\tfrac{1}{2}\pm\sqrt{(n+1)^2+g^2}
$$
have degeneracy $2n(n+2).$ The factor of $2$ arises because the energy
of these states is independent of parity. Furthermore, these states
only occur for $G>0.$

For zero coupling ($g=0$) the energy spectrum is $E=\pm(N+\tfrac{3}{2})$
and the degeneracy was found to be
\begin{equation}
D=4(N+1)(N+2),
\end{equation}
in agreement with \cite{Camporesi:1995fb}. We also found explicit
formulae for accidental degeneracies which occur for special values of
the coupling constant $g.$

The explicit formulae for the energy spectrum and its degeneracy
enabled us to write down the zeta function related to the Dirac sea.
For massive fermions on $S^3$, we were able to
derive an asymptotic formula for a zeta function of generalized
Epstein-Hurwitz type. The more interesting case of fermions coupled to
Skyrmions on $S^3$ leads to an interesting novel type of zeta function.
However, we were unable to evaluate it using our current
approach. This is an interesting topic for further study.

\section*{Acknowledgements}
The authors would like to thank N S Manton for fruitful discussions.
SWG would like to gratefully acknowledge funding from EPSRC and from
the School of Mathematics, Statistics and Actuarial Science at the
University of Kent.

\begin{small}

\end{small}
\end{document}